\documentclass[prb,twocolumn,amsmath,amssymb,floatfix]{revtex4}
\usepackage{graphicx}

\begin{document}

\title{Strong reduction of field-dependent microwave surface resistance in YBa$_{2}$Cu$_{3}$O$_{7-\delta}$ with sub-micrometric BaZrO$_3$ inclusions}

\author{N. Pompeo}
\affiliation{Dipartimento di Fisica ``E. Amaldi'' and Unit\`a CNISM, Universit\`a Roma Tre, Via della Vasca Navale 84, I-00146 Roma, Italy}
\author{R. Rogai}
\affiliation{Dipartimento di Fisica ``E. Amaldi'' and Unit\`a CNISM, Universit\`a Roma Tre, Via della Vasca Navale 84, I-00146 Roma, Italy}
\author{E. Silva\footnote{corresponding author. e-mail: silva@fis.uniroma3.it}}
\affiliation{Dipartimento di Fisica ``E. Amaldi'' and Unit\`a CNISM, Universit\`a Roma Tre, Via della Vasca Navale 84, I-00146 Roma, Italy}
\author{A. Augieri}
\affiliation{ENEA-Frascati, Via Enrico Fermi 45, 00044 Frascati, Roma, Italy}
\author{V. Galluzzi}
\affiliation{ENEA-Frascati, Via Enrico Fermi 45, 00044 Frascati, Roma, Italy}
\author{G. Celentano}
\affiliation{ENEA-Frascati, Via Enrico Fermi 45, 00044 Frascati, Roma, Italy}

\date{\today}

\begin{abstract}

We observe a strong reduction of the field induced thin film surface resistance measured at high microwave frequency ($\nu=$47.7 GHz) in YBa$_{2}$Cu$_{3}$O$_{7-\delta}$ thin films grown on SrTiO$_3$ substrates, as a consequence of the introduction of sub-micrometric BaZrO$_3$ particles. The field increase of the surface resistance is smaller by a factor of $\sim$3 in the film with BaZrO$_3$ inclusions, while the zero-field properties are not much
affected. Combining surface resistance and surface reactance data we conclude (a) that BaZrO$_3$ inclusions determine very deep and steep pinning wells and (b) that the pinning changes nature with respect to the pure film. 

\end{abstract}


\maketitle

Increasing pinning of vortex lines is an essential achievement for useful applications of superconductors in general, and of cuprate superconductors in particular due to the wide range of the $H-T$ phase diagram where vortices are free or nearly free to move. Columnar defects are best suited to this aim,\cite{civalePRL91} but the introduction of such defects requires rather sophisticated techniques which are not suitable for mass production. Recently it was shown \cite{macmanusNATMAT04} that inclusions of BaZrO$_3$ (BZO) particles of dimensions in the 10-1000 nm range could determine a significant improvement of the dc properties of YBa$_{2}$Cu$_{3}$O$_{7-\delta}$ (YBCO) thin films. In particular, the critical current density \cite{macmanusNATMAT04,kangSCI06} could be raised up to values useful for large scale dc applications, and the irreversibility line could be shifted above 10 T at 77 K.\cite{PeurlaPRB07} A major interest comes from the easy incorporation of such sub-$\mu$m particles directly into the targets used for film deposition, thus making feasible the mass growth of very low losses superconductors.\\
Pinning of vortices at high frequencies is desirable not only for devices operating in dc magnetic fields, but also because power handling is limited by depinning of self-nucleated vortices. In fact, tailored profiles of columnar defects have been shown to largely extend the linear regime at low microwave frequencies.\cite{ghigoSUST05} However, when the operating frequency increases, the dissipation due to oscillating vortices becomes increasingly difficult to reduce: with increasing frequency the amplitude of the vortex oscillation decreases, and becomes so small that only interactions between single vortices and pinning centers determine the response. In this regime the steepness of the pinning wells affects the balance between the reactive (elastic) and resistive (viscous) response, and the depth affects the creep of vortices. Even columnar defects, when the driving current oscillates in the high microwave range, determine at most a reduction of the dissipation of only $\sim$15\%, as seen from measurements at $\sim$50 GHz,\cite{silvaIJMPB00} while in some cases the microwave dissipation raised after heavy-ion irradiation.\cite{wosikAPL99}  It seems reasonable to state that, up to now, no known artificial pinning centers have shown a significant reduction of the dissipation at very high frequencies. Measuring the response at the high edge of the microwave spectrum is a very stringent test for the properties of artificial pinning centers.\\
In this Letter we report on the observation of a very large reduction of the field-induced surface resistance at 47.7 GHz in YBCO films as a consequence of the inclusion of sub-$\mu$m sized BZO particles. The increase of the surface resistance by the application of a dc magnetic field is smaller by a factor $\sim$3 in the film with BZO inclusions, up to very close to the transition temperature. By simultaneous measurements of the variation of the surface reactance we estimate the change of several vortex parameters. We argue that BZO inclusions act to substantially reduce both the creep of vortices (affected by the depth of pinning wells) and the mean vortex oscillations (affected by the steepness of pinning wells), which are particularly relevant features for high-frequency applications. From the field dependence of the pinning constant we argue that the nature of pinning changes from pinning of soft lines in the pure film to pinning of rigid, strongly correlated flux lines in YBCO/BZO.\\
Thin (thickness $d\approx$ 120 nm) YBCO films with $T_c\simeq$ 90 K were epitaxially grown on (001) SrTiO$_3$ (STO) substrates by high-oxygen pressure pulsed laser deposition technique using YBCO targets pure and additioned of 7 mol.\% BZO powder (YBCO/BZO). As estimated by SEM and optical microscope analysis, granularity of BZO powder in the target was below 1 $\mu$m. More details related to the BZO phase formation, composite target preparation as well as thin film deposition can be found elsewhere.\cite{galluzziIEEE07} X-ray diffraction of the YBCO/BZO sample revealed the existence of ($h$00)BZO reflections with intensity consistent with the amount of BZO in the target. Since the microscopic nature of the additional pinning centers related to the BZO incorporation is generally attributed both to edge dislocations nucleated at the YBCO/BZO interface and to the $c$-axis elongated character or self-alignment of BZO crystallites,\cite{macmanusNATMAT04,kangSCI06} we estimated the presence of linear defects in our YBCO/ BZO by repeated wet-chemical etching.\cite{huijbregtsePRB00} We found etch pits characteristic of linear defects parallel to the c-axis, with uniform distribution and areal density 30-40 times higher than in pristine YBCO. The film thickness was in the range were the growth technique yielded the most crystalline samples. Moreover, it allowed for a very reliable application of the thin film approximation in the analysis of the microwave data (see below). Several films have been examined, and consistently yielded higher $J_c$ in a magnetic field, as a consequence of the $c-$axis correlated nature of the pinning mechanisms active in films with BZO inclusions.\cite{galluzziIEEE07}  Two films with 0\% and 7\% have been selected for the study at microwave frequencies. BZO inclusions did not determine significant changes of the transport properties in zero field. In dc, $\rho(100 K)$ and $J_c(77K)$ changed from 125 $\mu \Omega$cm and 2.5 MA cm$^{-2}$ in pure YBCO to 110 $\mu\Omega$cm and 1.6 MA cm$^{-2}$, respectively, in YBCO/BZO at 7\%. The real microwave resistivity at 60 K and 47.7 GHz changed from 0.9 $\mu\Omega$cm to 1.2 $\mu\Omega$cm.\\
The field induced change in the intrinsic thin film impedance, $\Delta R'_{s}+\mathrm{i}\Delta X'_{s}=(\Delta \rho_{1}+\mathrm{i}\Delta \rho_2)/d$, was measured for $T>$ 60 K and $\mu_0H<$ 0.8 T by means of a sapphire dielectric resonator operating in the TE$_{011}$ mode at 47.7 GHz.\cite{pompeoJSUP07} The STO substrate resonances \cite{kleinJAP90,silvaSUST96} were removed with the introduction of an additional spacer to change the substrate impedance.\cite{pompeoPREP07} Here, $\rho_{1}+\mathrm{i} \rho_2$ is the complex resistivity of the film. Measurements were taken in the linear response regime, as ensured by calculations and direct verfication.\\
\begin{figure}[htbp]
\includegraphics[width=8cm]{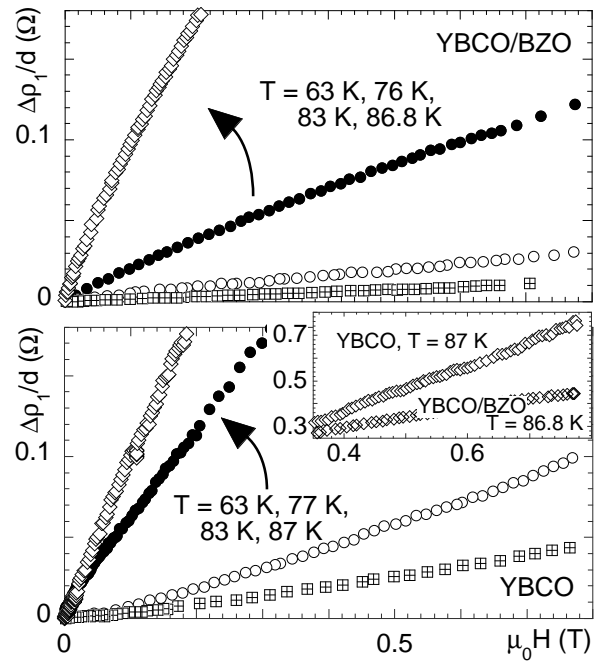}
\caption{Field dependence of the thin film surface resistance changes, $\Delta R_s'(H)=\Delta\rho_1(H)/d$ in YBCO/BZO (upper panel) and in pure YBCO (lower panel). In YBCO/BZO the dissipation is reduced by a factor $\sim$ 3 or larger. Even at 87 K, close to $T_c$, with increasing field the dissipation in YBCO/BZO is much lower than in YBCO (see inset).}
\label{fig1_RH}
\end{figure}
The main experimental finding of this work is exemplified in Fig.\ref{fig1_RH}, where we report $\Delta R_s'$ vs. $H$ at different temperatures. YBCO/BZO presents a dramatically reduced dissipation as a function of the applied field with respect to pure YBCO. Very interestingly, this reduction extends to temperatures close to $T_c$, indicating an exceptionally strong effect of the pinning centers arosen subsequently to the introduction of the BZO particles.
Figure \ref{fig1_RH} indicates that the introduction of BZO particles severely limits the maximum velocity (and as a consequence the oscillation amplitude) of the vortex lines.\\
The field induced changes in the reactance $\Delta X_s'$, reported in Fig. \ref{fig2_XH}, shows in YBCO/BZO a less dramatic reduction than the resistance changes at low $T$, but a strong persistence close to $T_c$, indicating that pinning remains strong at very high $T$. The significant reactance in pure YBCO indicates that even in the pure sample there were rather strong pinning centers (we recall our very high operating frequency, 47.7 GHz). Clearly, the introduction of BZO particles further enhance pinning at high frequency. We remind that BZO particles were found to further enhance the \textit{dc} properties of already strongly pinned samples.\cite{macmanusNATMAT04} In dc dissipation arises from the extraction of a vortex from a pinning center. It is a nontrivial finding that BZO inclusions originate pins that are effective at frequencies so high that shaking of vortices around their equilibrium position is the dominating mechanism.\\
\begin{figure}[htbp]
\includegraphics[width=8cm]{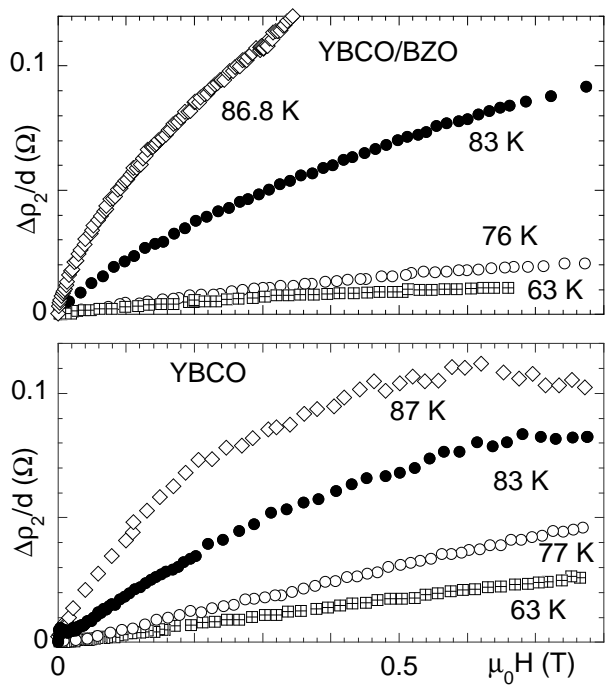}
\caption{
Field dependence of the thin film surface reactance changes, $\Delta X_s'(H)=\Delta\rho_2(H)/d$ in YBCO/BZO (upper panel) and in pure YBCO (lower panel). In YBCO/BZO the reactance remains large even at $T\approx T_c$, suggesting persistence of pinning.
}
\label{fig2_XH}
\end{figure}
Both $R_s'$ and $X_s'$ acquire in YBCO/BZO a downward curvature in their $H$ dependence. An \textit{upward} curvature in $R_s'(H)$ (as distinguishable in pure YBCO) is usually an indication of flux creep. Thus, our data suggest that introduction of BZO, in addition to strong pinning, leads to a reduction of creep at high $T$. We note that, in terms of the parameter $r=\Delta X_s'(H)/\Delta R_s'(H)$, customarily used to describe the balance between elastic (reactive) and viscous (resistive) vortex response,\cite{halbritter} YBCO/BZO results to be very strongly pinned, with $r\sim$2 at 64 K (Fig.\ref{fig3_param}).\\
To discuss more quantitatively our results we extract the vortex parameters from our data by means of the well-known Coffey-Clem (CC) model \cite{cc}:
\begin{equation}
    \Delta\rho_1+\mathrm{i}\Delta\rho_2=\frac{\Phi_0 B}{\eta}\frac{1+\epsilon\left(\frac{\nu_0}{\nu}\right)^2+ \mathrm{i} (1-\epsilon)\frac{\nu_0}{\nu}}{1+\left(\frac{\nu_0}{\nu}\right)^2}.
    \label{eqCC}
\end{equation}
where $\Phi_0$ is the flux quantum, the vortex viscosity $\eta$ is proportional to the density of states in the vortex core, $B\simeq \mu_0 H$ (London approximation), the creep factor 0 $\leq \epsilon \leq \epsilon_{max}=1+2r^2-2r\sqrt{r^2+1}$ is a measure of the height of the pinning potential ($\epsilon=$0 corresponds to no creep),\cite{epsmax} $\nu_0$ is a characteristic frequency and $\nu_o\rightarrow \nu_p$ (pinning frequency) for $\epsilon\rightarrow 0$, and the pinning constant $k_p=2\pi\nu_p\eta$ is a measure of the steepness of the pinning potential. Estimates of the maximum creep factor are reported in Fig. \ref{fig3_param}. While one should bear in mind that $\epsilon_{max}$ is only an upper limit for the creep factor, the differences between YBCO/BZO and YBCO are evident. In particular, $\epsilon$ does not exceed 0.4 in YBCO/BZO, even for $T\approx T_c$.\\
\begin{figure}
\includegraphics[width=8.5cm]{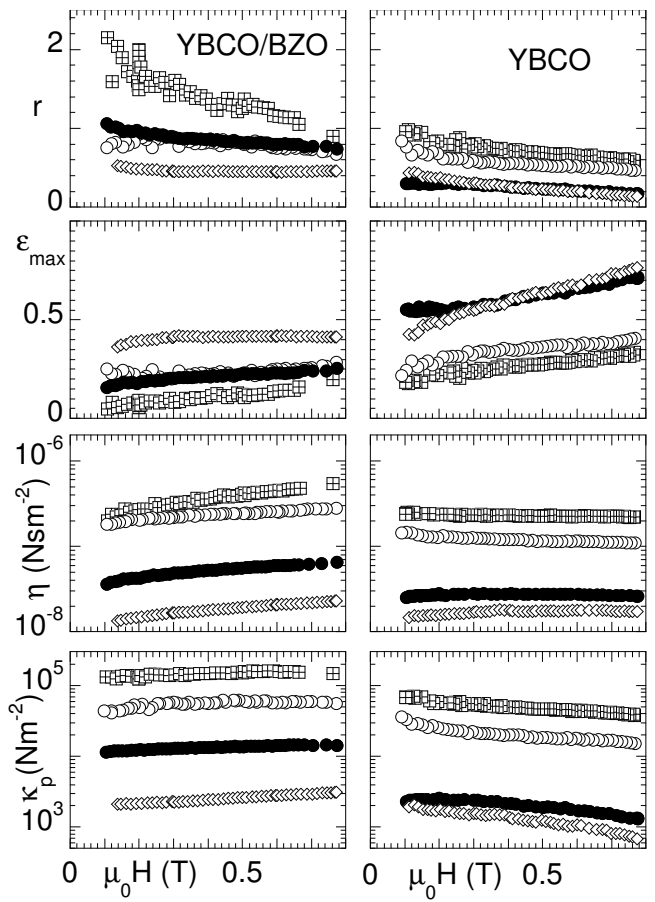}
\caption{
Vortex parameters at different temperatures. Symbols indicate same temperatures as in Fig.\ref{fig1_RH}. Left panels: YBCO/BZO, $T$=63, 76, 83, 86.8 K. Right panels: pure YBCO, $T$=63, 77, 83, 87 K. Note the persistent pinning constant and small maximum creep at 86.8 K in YBCO/STO.
}
\label{fig3_param}
\end{figure}
The remaining vortex parameters can be estimated using Eq.\ref{eqCC} varying $\epsilon$ from 0 to $\epsilon_{max}$, yielding estimates that, in our case, do not differ by more than 30\% for $\eta$ in the worst case. Estimates of $k_p$ are typically much more precise. We report the values calculated for $\epsilon=0$. The vortex viscosity attains typical values in this temperature range.\cite{golosovskySUST96} In YBCO/BZO $\eta$ increases with respect to YBCO, qualitatively consistent with $\rho_{dc}$ being smaller by $\sim$20\% as measured in similar samples, and shows a slight field dependence. This has been suggested to arise from the shrinking of vortex cores with the application of a magnetic field.\cite{koganCM} However, it is not clear at present why this effect should be absent (or non detectable) in pure YBCO. We leave this topic, relevant for the basic aspects of the vortex physics, to a future investigation, and we focus here on the pinning properties, exemplified by the field dependence of the pinning constant $k_p$. The absolute values of $k_p$ in YBCO are aligned with typical values\cite{golosovskySUST96} obtained from radiofrequency and microwave measurements, while in YBCO/BZO an enhancement takes place in particular around the liquid nitrogen temperature. Moreover, and most important, we find that in YBCO $k_p$ is a decreasing function of $H$: this feature is typical of collective pinning, arising from fluxon softening.\cite{golosovskySUST96}  By contrast, in YBCO/BZO we find $k_p\sim const$ or slightly increasing with $H$, indicating that flux lines act essentially as rigid rods.\\
Summarizing, we have presented experimental evidence of strong pinning in YBCO/BZO with respect to small oscillations induced by microwave currents. We have found strong reduction of the field-induced dissipation. By extracting the vortex parameters we have argued that BZO particles induce very steep and deep pinning wells. The vortex creep is reduced by BZO particles. The pinning constant increases and remains large at high $T$. The combination of these effects make the addition of BZO particles particularly useful in order to improve the high-frequency performances of YBCO.\\
\\
We are grateful to T. Petrisor and L. Ciontea for target preparation.



\begin{thebibliography}{99}

%
\bibitem{civalePRL91} L. Civale, A. D. Marwick, T. K. Worthington, M. A. Kirk, J. R. Thompson, L. Krusin-Elbaum, Y. Sun, J. R. Clem, F. Holtzberg, {\it Phys. Rev. Lett.} {\bf 67}, 648 (1991).

%
\bibitem{macmanusNATMAT04} J. L. Macmanus-Driscoll, S. R. Foltyn, Q. X. Jia, H. Wang, A. Serquis, L. Civale, B. Malorov, M. E. Hawley, M. P. Maley, D. E. Peterson, {\it Nat. Mater.} {\bf 3}, 439 (2004).

%
\bibitem{kangSCI06} S. Kang, A. Goyal, J. Li, A. A. Gapud, P. M. Martin, L. Heatherly, J. R. Thompson, D. K. Christen, F. A. List, M. Paranthaman and D. F. Lee, {\it Science} {\bf 311}, 1911 (2006).

%
\bibitem{PeurlaPRB07}  M. Peurla, H. Huhtinen, M. A. Shakhov, K. Traito, Yu. P. Stepanov, M. Safonchik, P.Paturi, Y. Y. Tse, R. Palai, R. Laiho,  {\it Phys. Rev. B} {\bf 75}, 184524 (2007).

%
\bibitem{ghigoSUST05}  G. Ghigo, D. Andreone, D. Botta, A. Chiodoni, R. Gerbaldo, L. Gozzelino, F. Laviano, B. Minetti and E. Mezzetti, {\it Supercond. Sci. Technol.} {\bf 18}, 193 (2005).

%
\bibitem{silvaIJMPB00} E. Silva, G. Ghigo, L. Gozzelino, C. Camerlingo, S. Sarti, {\it Int. J. of Mod. Phys. B} {\bf 14}, 2822 (2000).
%
\bibitem{wosikAPL99} J. Wosik, L.-M. Xie, J. Mazierska and R. Grabovickic, {\it Appl.Phys. Lett.} {\bf 75}, 1781 (1999).

%
\bibitem{galluzziIEEE07} V. Galluzzi, A. Augieri, L. Ciontea, G. Celentano, F. Fabbri, U. Gambardella, A. Mancini, T. Petrisor, N. Pompeo, A. Rufoloni, E. Silva, A. Vannozzi, {\it IEEE Trans. Appl. Supercond.} {\bf 17}, 3628 (2007)

\bibitem{huijbregtsePRB00} J.M. Huijbregtse, B. Dam, R. C. F. van der Geest, F. C. Klaassen, R. Elberse, J. H. Rector, and R. Griessen, {\it Phys. Rev. B} {\bf 62}, 1338 (2000).

\bibitem{pompeoJSUP07} N. Pompeo, R. Marcon and E. Silva, {\it 
J. Supercond. and Novel Magnetism} {\bf 20}, 71 (2007).

%
\bibitem{kleinJAP90} N. Klein, H. Chaloupka, G. M\"{u}ller, S. Orbach, H. Piel, B. Roas, L. Schultz, U. Klein and M. Peiniger, {\it J. Appl. Phys.} {\bf 67}, 6940 (1990).

\bibitem{silvaSUST96} E. Silva, M. Lanucara and R. Marcon,  {\it 
Supercond. Sci. Technol.} {\bf 9}, 934 (1996).

\bibitem{pompeoPREP07} N. Pompeo, L. Muzzi, V. Galluzzi, R. Marcon and E. Silva, {\it Supercond. Sci. Technol.} {\bf 20}, 1002 (2007).

\bibitem{cc} M. W. Coffey and J. R. Clem, {\it Phys. Rev. Lett.} {\bf 67}, 386 (1991).


\bibitem{halbritter} J. Halbritter, {\it J. Supercond.} {\bf 8} 691 (1995).

\bibitem{epsmax} $\epsilon_{max}$ is an analytical property of the CC model, N. Pompeo, R.Marcon, S. Sarti, H. Schneidewind, E. Silva,  {\it J. Supercond. and Novel Magnetism} {\bf 20}, 43 (2007).
%

\bibitem{golosovskySUST96} M. Golosovsky, M. Tsindlekht and D. Davidov, {\it Supercond. Sci. Technol.} {\bf 9}, 1 (1996).
%

\bibitem{koganCM} V. G. Kogan, R. Prozorov, S. L. Bud'ko, P. C. Canfield, J. R. Thompson, J. Karpinski, N. D. Zhigadlo, P. Miranovi\'c, {\it cond-mat/0606562} (2006).

\end{thebibliography}
\end{document}